\documentclass[preprint2]{aastex}
\usepackage{natbib}

\def\apjs{ApJS}

\def\apjl{ApJ}
\def\aap{A\&A}

\def\asec{\ifmmode ^{\prime\prime}\else$^{\prime\prime}$\fi}

\def\it{\sl}
\def\degs{\ifmmode ^{\circ}\else$^{\circ}$\fi}
\def\amin{\ifmmode ^{\prime}\else$^{\prime}$\fi}
\def\asec{\ifmmode ^{\prime\prime}\else$^{\prime\prime}$\fi}
\def\fdg{\hbox{$.\!\!^\circ$}}          
\def\farcs{\hbox{$.\!\!^{\prime\prime}$}}  

\def\degs{\ifmmode ^{\circ}\else$^{\circ}$\fi}
\def\amin{\ifmmode ^{\prime}\else$^{\prime}$\fi}
\def\farcm{\hbox{$.\mkern-4mu^\prime$}}
\def\eqalign#1{\null\,\vcenter{\openup1\jot \m@th
   \ialign{\strut\hfil$\displaystyle{##}$&$\displaystyle{{}##}$\hfil
   \crcr#1\crcr}}\,}

\slugcomment{to appear in Astrophysical J., 45.}

\shorttitle{Thermal properties of the $\gamma$-ray pulsar J1741$-$2054}
\shortauthors{Karpova et al.}

\begin{document}


\title{Thermal properties of the middle-aged pulsar J1741$-$2054
    }


\author{A.~Karpova\altaffilmark{1}, A.~Danilenko, Yu.~Shibanov\altaffilmark{1}, 
P.~Shternin\altaffilmark{1}, and D.~Zyuzin}
\affil{Ioffe Physical Technical Institute, Politekhnicheskaya
 26, St. Petersburg, 194021, Russia}


\altaffiltext{1}{St.~Petersburg State Polytechnical University, 
Politekhnicheskaya 29, St. Petersburg, 195251, Russia}


\begin{abstract}    
  We present results of the spectral analysis of the X-ray emission from
  the middle-aged \textit{Fermi} pulsar J1741$-$2054 using all
  \textit{Chandra} archival data collected in 2010 and 2013. We confirm early findings by
  Romani et al. 2010 that the pulsar spectrum contains a thermal emission
  component. The component is best described by the blackbody model with the
  temperature $\approx$~60 eV
  and the emitting area radius $\approx$~17~$D_{\rm kpc}$ km. 
  The thermal emission 
  likely originates
  from the entire surface of the cooling neutron star if the distance to the pulsar is
  $\approx$~0.8 kpc.
  The latter is supported by a large absorbing column density
  inferred from the X-ray fit and empirical optical
  extinction--distance relations along the pulsar line of sight.
  The neutron star surface temperature and characteristic age make 
  it similar to the well studied middle-aged pulsar B1055$-$52.
  Like the latter PSR J1741$-$2054 is hotter than the standard cooling
  scenario predicts.
\end{abstract}

\keywords{
stars: neutron -- pulsars: general -- pulsars: individual: PSR J1741$-$2054.
}

\section{Introduction}

One of a few possibilities to probe the physics of the dense 
matter in extreme conditions inside 
neutron stars (NSs) 
is to study the thermal emission from their surfaces \citep{hpyBOOK}. 
Nearby middle-aged pulsars are natural targets for these aims. 
Many of them clearly show the thermal component in their 
soft X-ray spectra. In some cases it can be attributed 
to the emission from the entire NS surface and  
the surface temperature can be measured. So far it 
has been done only for a dozen of middle-aged pulsars.
Recent \textit{Fermi-LAT} $\gamma$-ray discoveries 
open a new window for that.

A middle-aged $\gamma$-ray pulsar J1741$-$2054, hereafter J1741,
has a period $P$~= 413~ms, a characteristic age $\tau_c$~= 391~kyr,
a spin-down luminosity $\dot{E}$~= 9.5~$\times$~10$^{33}$ erg~s$^{-1}$, and
a magnetic field $B$~= 2.7~$\times$~10$^{12}$~G \citep{abdo2013ApJS}. 
After the discovery with \textit{Fermi} \citep{abdo2009} it was studied in various bands. 
A dispersion measure (DM) of 4.7~pc~cm$^{-3}$ \citep{camilo2009ApJ} implies
a distance  of 
380 pc for the Galactic
electron density model of \citet{cordes2002astro.ph}. 
A bow-shock nebula 
was detected around the pulsar in H$\alpha$ \citep{romani2010ApJ}.
In X-rays, a point-like object identified with the pulsar, a compact 
pulsar wind nebula (PWN) structure, and a PWN trail
extended through 7\arcsec~and 2\arcmin~from the pulsar, 
respectively, were detected with \textit{Chandra} \citep{romani2010ApJ}.

\citet{romani2010ApJ} pointed out the presence of a soft 
thermal component in the pulsar spectrum, however
no analysis of the thermal emission have been presented so far,
save for a brief conference abstract by \citet{sivakoff2011HEAD}.
Here we fill this gap and report results of 
analysis of the pulsar thermal emission.

\section[]{Analysis of the X-ray spectrum}
\begin{table}[tbh]
\caption{Chandra observations of J1741.}
\label{t:ch_obs}
\begin{center}
\begin{tabular}{cccccccc}
  \tableline\tableline
  Obs ID & Instrument & Exp., ks & Start Date \\ \hline
  11251  & ACIS-S     & 48.78        & 2010-05-21 \\
  14695  & ACIS-S     & 57.15        & 2013-02-06 \\
  14696  & ACIS-S     & 54.3         & 2013-02-19 \\
  15542  & ACIS-S     & 28.29        & 2013-04-01 \\
  15543  & ACIS-S     & 57.22        & 2013-05-17 \\
  15544  & ACIS-S     & 55.73        & 2013-07-12 \\
  15638  & ACIS-S     & 29.36        & 2013-04-02 \\ \hline
\end{tabular}
\end{center}  
\end{table}

\begin{figure*}[t]
\setlength{\unitlength}{1mm}
  \begin{center}
    \begin{picture}(145,60)(0,0)
 \put (79,10) {\includegraphics[scale=0.47, clip]{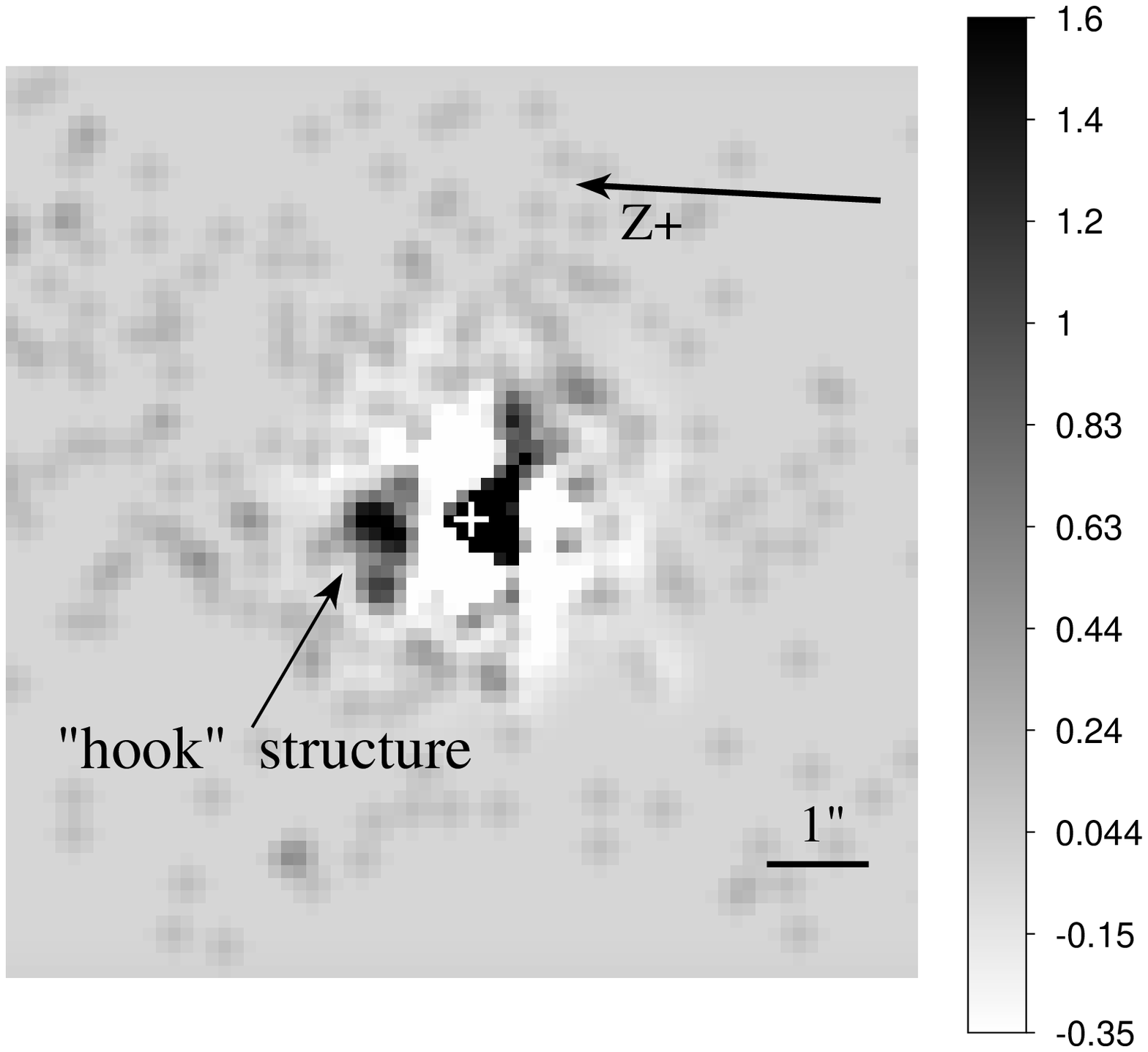}}
 \put  (-10,0){\includegraphics[scale=0.5, clip]{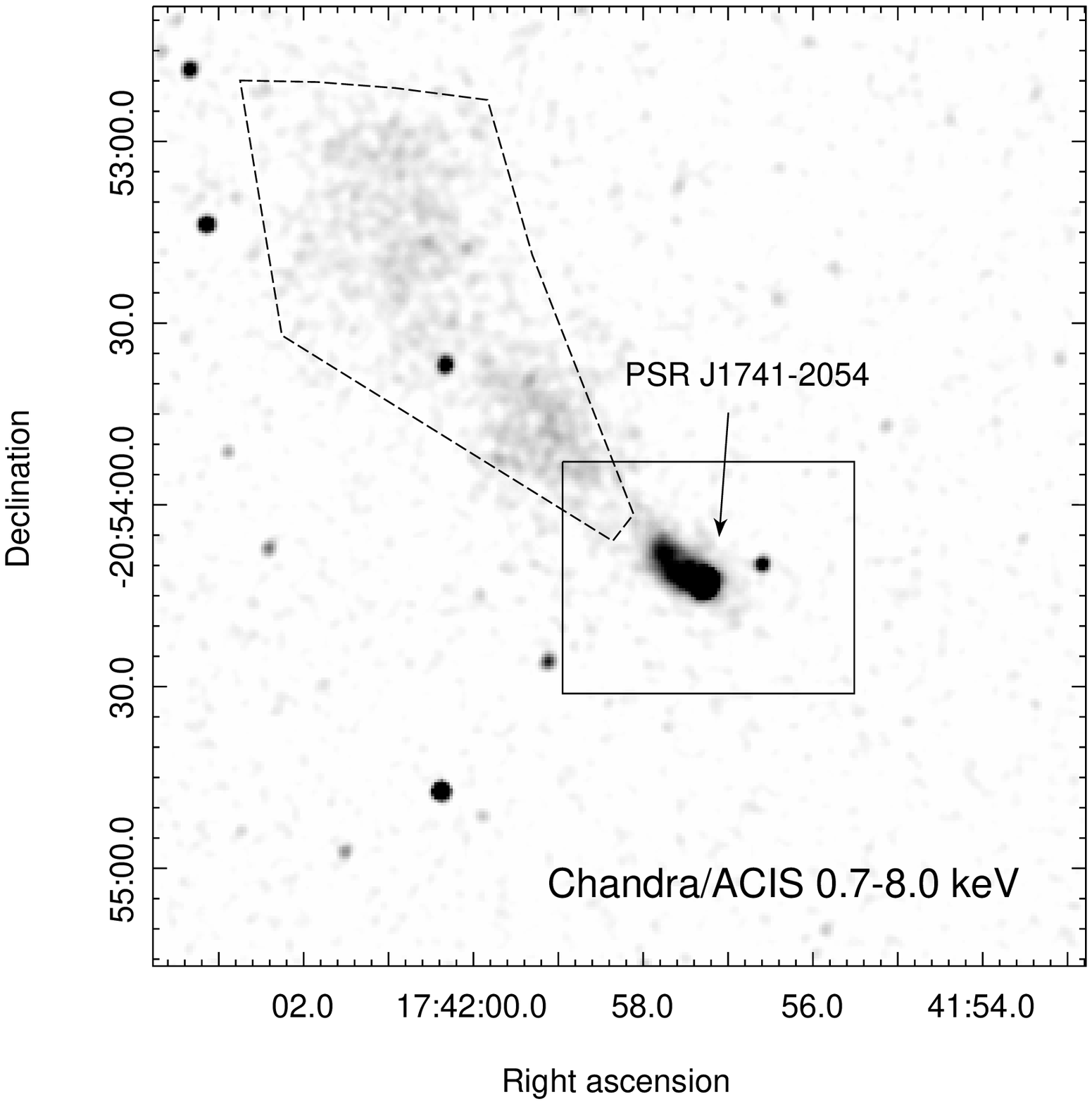}} 
 \put (48.5,64.5) {\includegraphics[scale=0.15, clip]{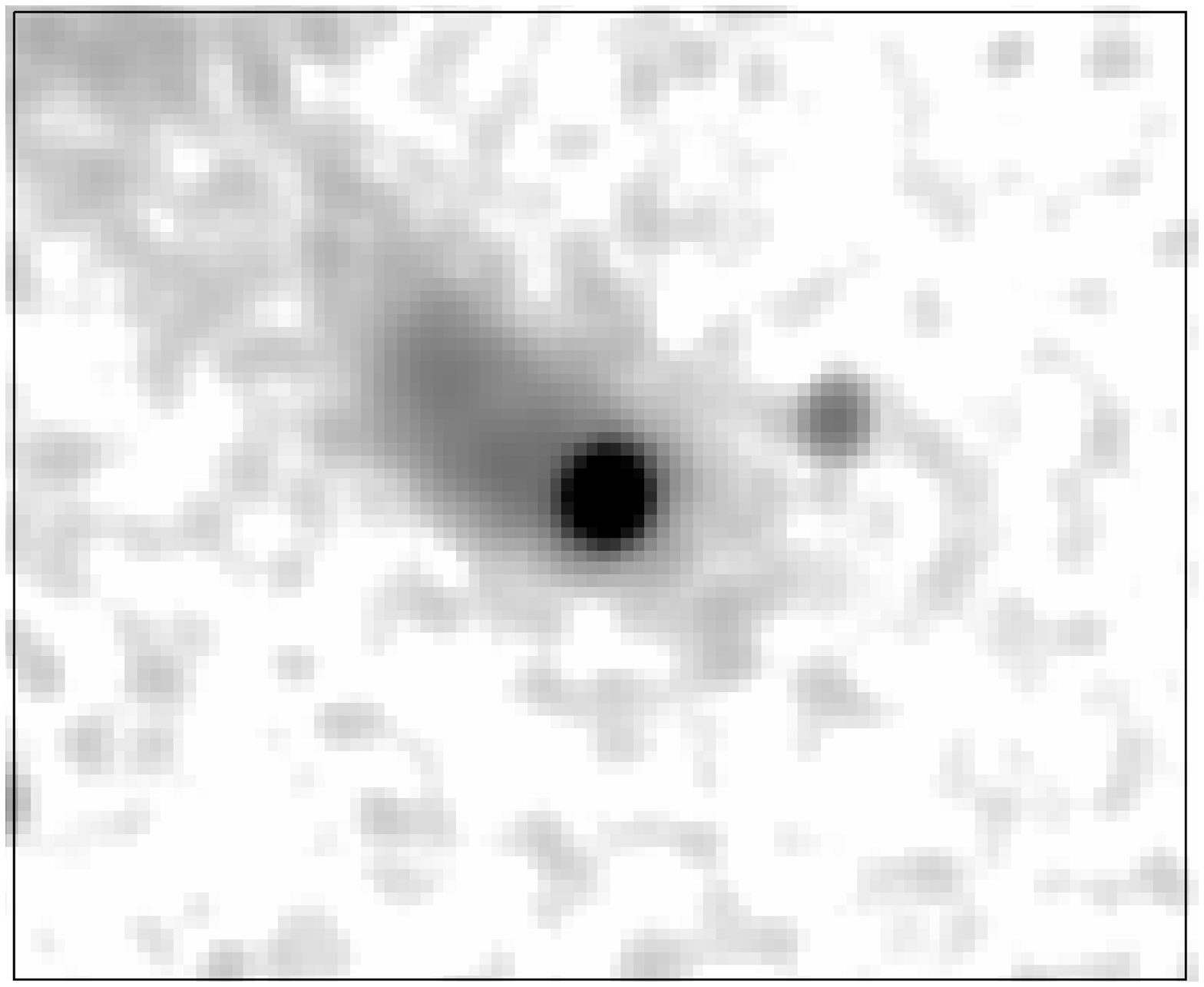}}
 \end{picture}
  \end{center}
  \caption{{\sl Left:} 2\farcm6~$\times$~2\farcm6 fragment 
  of the pulsar field in 0.7--8.0 keV as seen with \textit{Chandra}/ACIS, 
  smoothed with a three-pixel Gaussian kernel. The 48\farcs8~$\times$~40\farcs3 
  pulsar vicinity enclosed in the black box  
  is enlarged in the inset, where logarithmic brightness scale is used. 
  The dashed polygon encloses the PWN trail.
  {\sl Right:} PSF fit residuals for ObsID 11251. Cross marks the 
  pulsar position. The $Z+$ detector axis direction and the ``hook'' structure are indicated.}
\label{fig:merged}
\end{figure*}

\begin{figure}[t]
\setlength{\unitlength}{1mm}
  \begin{center}
    \begin{picture}(145,61)(0,0)
 \put (0,0) {\includegraphics[scale=0.4, clip]{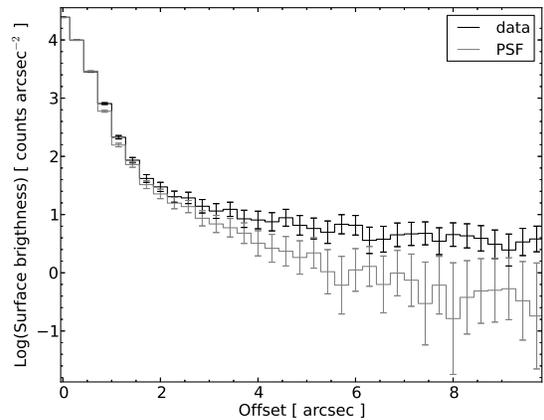}}
 \end{picture}
  \end{center}
  \caption{Comparison of the observed radial brightness 
  profile with the simulated PSF in the pulsar vicinity.}
\label{fig:profile}
\end{figure}

Recently, the J1741 field has been extensively observed in X-rays with
\textit{Chandra}/ACIS as part 
of the large program\footnote{``A Legacy Study of the Relativistic Shocks of PWNe'',
PN 14500153, PI Roger Romani.}  
which aimed at studies of PWNe spatial and temporal variability. 
As a spin-off of the program, a substantial number of 
the pulsar photons has been collected.
We have retrieved all the data  
obtained with \textit{Chandra} 
(see Table~\ref{t:ch_obs}).
For all data sets, the data mode was VFAINT, the exposure mode was TE,
and the pulsar was exposed on the S3 chip.
The \texttt{CIAO v.4.6}  \texttt{chandra\_repro} 
tool with \texttt{CALDB v.4.5.9} was used to reprocess all data sets.

An image of the pulsar neighborhood in 0.7--8.0 keV 
obtained by merging all data sets
is shown in the left panel of Fig.~\ref{fig:merged}, where 
the pulsar
is marked by an arrow.
The compact PWN adjacent to the pulsar and the extended trail, 
enclosed in a dashed polygon, are
clearly seen. Based on the analysis of the ObsID 11251, \citet{romani2010ApJ}
reported on a tentative detection of a presumed torus of the PWN within the 2\arcsec\ from the pulsar
containing about 11\% of the total data counts in this region.
This conclusion followed from the appearance of the 2\farcs5~$\times$~0\farcs75 excess structure 
centered at the pulsar position after subtraction of the point-spread function (PSF) 
modeled by \texttt{MARX}. 
However, it is well known that the \textit{Chandra} PSF does not fit well
to the point-source core and that there 
are some anisotropic irregularities 
in the fit residuals.\footnote{http://cxc.harvard.edu/cal/Hrc/PSF/acis\_psf\_2010oct.html}
The \textit{Chandra} team advices to treat with caution
features with  small spatial dimensions and strengths 
like the presumed torus.\footnote{http://cxc.harvard.edu/ciao/caveats/psf\_artifact.html\#advice}

To examine how the point-like pulsar is actually contaminated by the
compact PWN we modeled 
PSF event files for each data set 
using the \texttt{ChaRT} \citep{carter2003ASPC} and \texttt{MARX} tools.
We have estimated 
the number of residual counts with respect to that of the point-like source
as follows.\footnote{for details see the corresponding \texttt{ciao} thread,
http://cxc.harvard.edu/sherpa/threads/2dpsf/} 
The data images were fitted with a sum of a symmetric 2D Gaussian, which modeled 
a point-like source blurring due 
to pointing uncertainty \citep[see e.g.,][]{weisskopf2011arXiv},
and a constant 
background, convolved with 
the modeled PSFs.  
The 2D residuals pattern is similar to those reported by 
\textit{Chandra} team for a number of point sources. The example for ObsID 11251 
is shown in the right panel of Fig.~\ref{fig:merged}.
We do not see evidence of any PWN structure in the central 2\arcsec.
The most prominent south-east structure correlated with the $Z+$ detector 
axis and 
containing $\approx 3 \%$ of the total source counts
is consistent with a ``hook feature'' described in the \textit{Chandra} report.
It is possibly related to imperfect modeling of the \textit{Chandra} optics. 
The putative PWN torus claimed by \citet{romani2010ApJ} could 
have been actually mixed up with this hook structure; in any case 
its detection is likely a result of an 
incorrect treatment of the systematic effects of the PSF subtraction. 

\begin{figure}[t]
\setlength{\unitlength}{1mm}
    \begin{picture}(145,91)(0,0)
\put (-3.0,0) {\includegraphics[scale=0.43, clip]{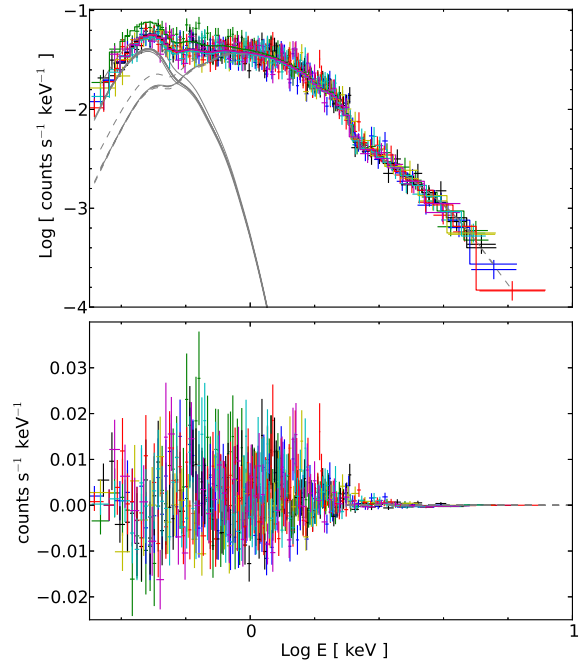}}
\end{picture}
  \caption{Folded best-fit BB+PL model,
  the data points ({\sl top}), and the fit residuals ({\sl bottom}).
  The folded BB and PL components are shown separately for each data set 
  by solid and dashed gray lines, respectively.
  }
\label{fig:spec}
\end{figure}

In order to select the reliable aperture for extraction of the pulsar spectra we 
calculated brightness profiles from the PSFs and data event files 
using concentric annuli centered at the pulsar  
position with radii from 0 to 10\arcsec~through 0\farcs29
and summed up the data and PSF profiles over all data sets. The resulting profiles are shown 
in Fig.~\ref{fig:profile}. 
It is seen that the data are dominated by the point source within the central
2\arcsec. 
We conservatively chose the 1\farcs5-radius  
aperture which contains
$\gtrsim$~95\% of the source 
counts. 
For the background, 
we used a region free 
of any sources in each data set. 
We extracted spectra with 
the \texttt{CIAO} \texttt{v.4.6} {\sl specextract } tools.
The spectra were grouped to ensure $\gtrsim$ 25 counts per energy bin.
Using the \texttt{XSPEC} \texttt{v.12.8.1} 
we then fitted the spectra in 0.3--10 keV range simultaneously by 
an absorbed composite model, a sum of a power law (PL), 
which models
the pulsar magnetosphere emission, and a thermal 
component, which models emission from 
the NS surface. For the thermal 
component, we have tried magnetized neutron star atmosphere models 
NSA \citep{pavlov1995lns} and NSMAX \citep{ho2008ApJS}, 
and the blackbody (BB) model. 

The goodness-of-fit test shows that 
the hydrogen atmosphere models
can be rejected 
for any reasonable model parameters. The reduced chi-squared value is
$\chi^{2}_{\nu}=1.41$ (dof~$=557$) for the NSA+PL and NSMAX+PL models.
The single BB and PL, and  BB+BB models give even worse fits:  $\chi^{2}_{\nu}=8.09$   
(dof~$=559$), $\chi^{2}_{\nu}$ = 1.91 (dof~$=559$),  and 
$\chi^{2}_{\nu}=1.73$ (dof~$=557$), respectively.

In contrast, the BB+PL model is statistically acceptable {\bf ($\chi^{2}_{\nu}=1.06$,  dof~$=557$)}.
It is compared with the data in Fig.~2 and its best-fit parameters are presented in Table 2. 
At the DM distance, the emitting area radius inferred from the fit
is too large for a hot polar cap, which  
should be about 200 m for a 400 ms 
pulsar according to \citet{sturrock1971ApJ}. At the same time, rather large 
emitting area radius suggests that the thermal emission originates
from a substantial part of the NS surface.

\begin{table}[t]
\caption{The absorbed BB+PL best-fit spectral parameters. All errors are at 90\% confidence.}
\label{t:best-fit}
\begin{center}
\begin{tabular}{ll}
\tableline
\multicolumn{2}{c}{ }  \\
absorption                        &  \\
 column density $N_{\rm H}$       &  $1.38^{+0.19}_{-0.19}$~$\times$~10$^{21}$ cm$^{-2}$                                                   \\  
\multicolumn{2}{c}{ }  \\
photon index $\Gamma$             & $2.66^{+0.06}_{-0.06}$                           \\ 
\multicolumn{2}{c}{ }  \\
PL normalization                  & $1.2^{+0.06}_{-0.06}$~$\times$~10$^{-4}$ \\
                                  & ph keV$^{-1}$~cm$^{-2}$~s$^{-1}$  \\
\multicolumn{2}{c}{ }  \\
emitting area radius $R$          & $17^{+3.5}_{-3}$ $D_{\rm kpc}$ km \\
\multicolumn{2}{c}{ }  \\
temperature   $T$                 & $60.0^{+2}_{-2}$ eV   \\ 
\multicolumn{2}{c}{ }  \\
$\chi^2_{\nu}$ (dof)              & 1.06 (557)           \\
 \multicolumn{2}{c}{ }  \\
 \hline
\end{tabular}
\end{center}
\end{table}

Fitting the PWN trail spectrum extracted from the dashed polygon
in Fig.~\ref{fig:merged} with the absorbed PL model we
get $N_{\rm H}$~= (1.6~$\pm$~0.5)~$\times$~10$^{21}$~cm$^{-2}$ and
photon index $\Gamma$~$=$ 1.78~$\pm$~0.15 (errors are at 90\%), which is in agreement with
the results of \citet{romani2010ApJ} and with the
$N_{\rm H}$ inferred from the BB+PL fit (Table~\ref{t:best-fit}). 
An independent 
limit on the absorption column density, based on the H$\alpha$ nebula spectroscopy, 
is $N_{\rm H}$~$<$ 2.5~$\times$~10$^{21}$~cm$^{-2}$ \citep{romani2010ApJ}.
To reconcile the large $N_{\rm H}$ values followed 
from both the X-ray and the H$\alpha$ nebula spectroscopy
with the electron column density 
$N_e$~$\approx$~1.4~$\times$~10$^{19}$ cm$^{-2}$ 
deduced from
the DM one needs to assume a low ionization 
ratio of about 0.01 along the line of sight \citep{romani2010ApJ}.

\begin{figure*}[th]
  \begin{center}
  \setlength{\unitlength}{1mm}
  \resizebox{15.5cm}{!}{
  \begin{picture}(150,43.5)(0,0)
    \put (-10.0,0) {\includegraphics[scale=0.42]{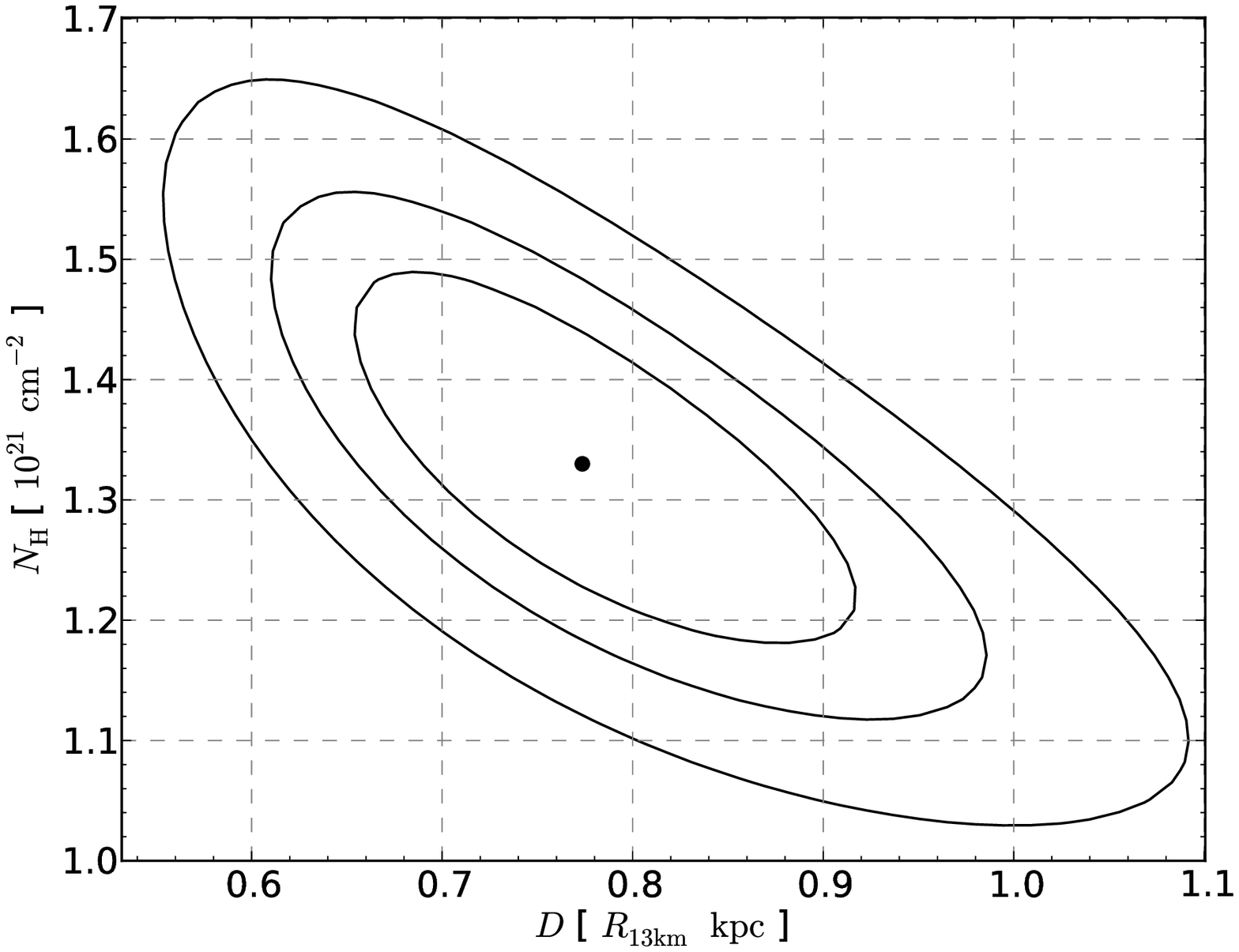}}
    \put (72,0) {\includegraphics[scale=0.42]{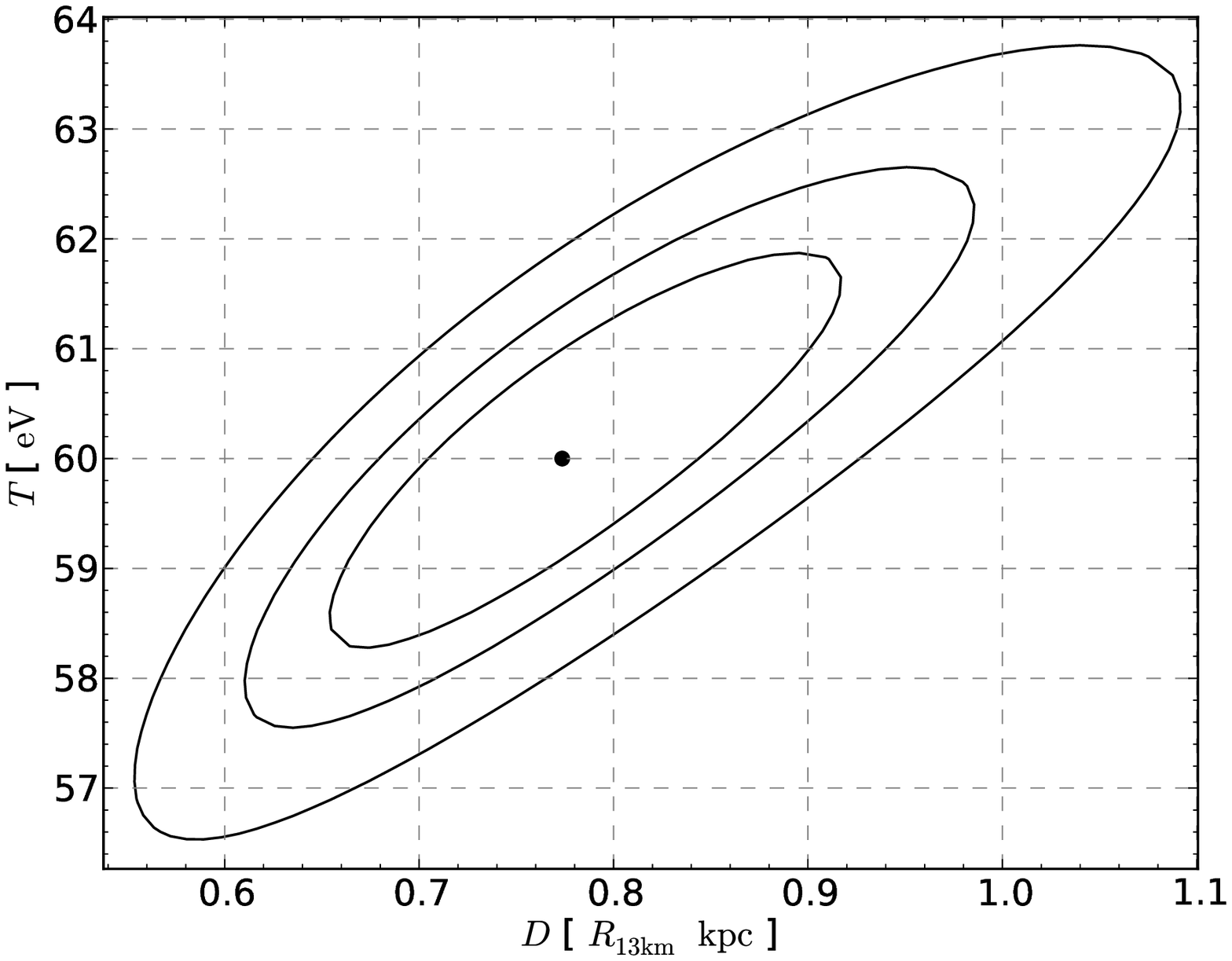}}
  \end{picture}
  }
  \end{center}
  \caption{
  68\%, 90\%, and 99\% 
  ($\Delta\chi^{2}=\chi^{2}-\chi^{2}_{min}$ = 2.3, 4.61, and 9.21, respectively) 
  confidence contours
  of the neutral hydrogen column density ({\sl left}) and the effective temperature 
  ({\sl right}) vs. the distance to the pulsar for the BB+PL model.  
  }
  \label{fig:contours}
\end{figure*}

\section{Discussion}

It is possible that the thermal component of the NS X-ray emission 
originates from the entire surface of the NS. 
The apparent radius of an NS with canonical parameters 
($M_{\rm NS}=1.4 M_\odot$ and $R_{\rm NS}=10$~km) is approximately 13~km.
In Fig.~\ref{fig:contours}, we 
show the confidence contours of the effective temperature 
$T$ and the absorbing column density $N_{\rm H}$ vs.
the distance to the pulsar $D$ evaluated from the BB normalization.
One sees that the pulsar should 
be at (0.8~$\pm$~0.2) $R_{\rm 13km}$~kpc which is a factor of two larger 
than the DM distance. 

We may estimate the distance independently as follows. 
The $N_{\rm H}$ value derived from the X-ray fit is nearly half   
the entire Galactic $N_{\rm H}$
in the pulsar direction, 3~$\times$~10$^{21}$~cm$^{-2}$, 
according to the HI survey by 
\citet{dickey1990ARAA}.
Assuming the Galactic gaseous disk half-thickness
of $\sim$~0.1 kpc and taking the pulsar latitude 
$b$~= 4\fdg9 
the maximal distance inside the
Galactic disk in the pulsar direction 
is about 1.4 kpc. Making a rough assumption 
that $N_{\rm H}$ 
is proportional to the distance, 
we conclude that
the pulsar must be approximately at half the maximal distance, 
which is 0.7 kpc. One of precautions this estimate should 
be taken with is that $N_{\rm H}$ values
from HI surveys may underestimate a column density 
responsible for the X-ray absorption.
However, the similar value of $0.7$~kpc 
is obtained from the optical extinction ($A_V$) --
distance fit in the pulsar direction \citep{chen1998AsAp},
where $A_V\approx 0.76$ is derived from the 
standard $A_V-N_H$ relation \citep{predehl1995AsAp}. 
Taking another $A_V$--distance fit from \citet{drimmel2003AsAp} we get
the distance of 0.9 kpc.
While the use of the extinction maps can underestimate 
the value along the specific line of sight due to the
lack of high  spatial resolution information, these distance estimates  
are in agreement with that 
obtained from the X-ray fit
(see Fig.~\ref{fig:contours}).
For completeness, the distance can be also estimated 
from an empirical correlation between the pulsar distances and $\gamma$-ray fluxes above 100~GeV. 
This ``pseudo-distance'' relation \citep[e.g.,][]{sazparkinson2010ApJ} 
suggests a value $\sim$450 pc. However this is the most uncertain estimate (within a factor of 2-3) and thus
is consistent with all other.


An increase in the distance 
means also an increase in the 
X-ray and $\gamma$-ray efficiencies, i.e. ratios of the 
X-ray nonthermal 
and $\gamma$-ray
luminosities to $\dot{E}$. The X-ray efficiency in 
the 2--10 keV range  derived from the X-ray fit is 
9.7~$\times$~10$^{-5}$~$D_{\rm 0.8kpc}^{2}$. 
The $\gamma$-ray efficiency is 0.97~$D_{\rm 0.8kpc}^{2}$
\citep{abdo2013ApJS}.
Such a high $\gamma$-ray efficiency is not unusual. There are $\gamma$-ray 
pulsars, Geminga for instance, with efficiencies almost equal or even greater than
one \citep[][]{abdo2013ApJS}.

\label{sec4}
\begin{figure}[t]
  \begin{center}
   \includegraphics[scale=0.41]{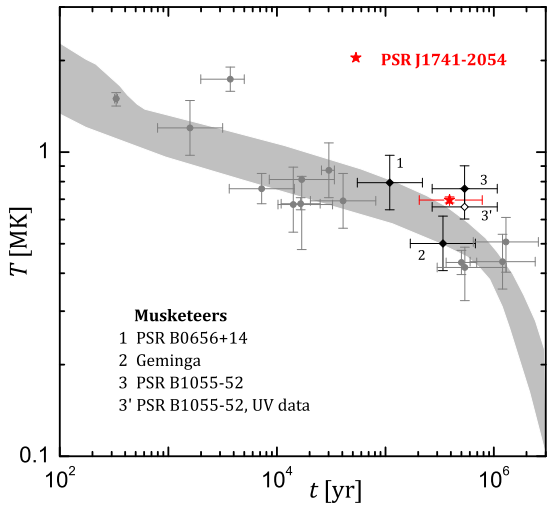}
  \end{center}
  \caption{
  Observations of isolated cooling NSs compared with standard cooling theory
  predictions (filled region).
  The PSR J1741$-$2054 data point is shown by the star symbol. We artificially
  adopt a factor of 2 error on its age.
  Three Musketeers are indicated with the filled diamonds. The
  temperature and age error-bars for these stars are shown in accordance with
  \citet{kaminker2006MNRAS} and \citet{yakovlev2008AIPC}. The open
  diamond shows the PSR B1055$-$52 position as follows from the UV data,
  see text for details.
  }
\label{fig:cool}
\end{figure}

It is natural to compare J1741
with ``The Three Musketeers'', Geminga, PSR B0656+14, and PSR B1055$-$52, 
well-studied middle-aged pulsars with 
observed thermal emission
 \citep{becker1997AsAp}.
Like the Three Musketeers, 
J1741 demonstrates a soft thermal emission in 
X-rays, which is well fitted by the blackbody model,
while all hydrogen atmosphere models fail giving
too large normalizations. In contrast to 
the Three Musketeers, however, the J1741 spectrum does 
not show an additional hot BB component.
The thermal component of J1741 is not as prominent with respect 
to the nonthermal one as the thermal components of the Three Musketeers.
This can explain the lack of a hot spot component in the spectral fit -- it is possibly 
hidden under the strong nonthermal component.
It is worth noting that 
J1741 has a relatively large power-law
photon index  ($\Gamma\sim 2.7$) in comparison with the musketeers ($\Gamma\lesssim 2$).

Let us compare the thermal properties of 
J1741 with data on other isolated NSs.
In Fig.~\ref{fig:cool}, we show
the standard NS cooling theory predictions
\citep[filled region; e.g.,][]{yakovlev2004ARA} along with the
observational data on the age--temperature plane. 
The data are taken from the same references
as cited in \citet{shternin2011MNRAS} with addition of the data on
PSR J0357+0325 from \citet{kirichenko2014AsAp}. The three
musketeers are marked by black diamonds and 
numbered. The J1741 position is shown with a star. 
From the point of view of the NS cooling theory 
J1741 is most similar to PSR B1055$-$52.  Like the latter,
it is located at the ``knee'' of the cooling curve, which
corresponds to the transition from the neutrino cooling stage to
the photon cooling stage \citep{yakovlev2004ARA}. 
One sees that J1741 is somewhat older and hotter than the standard
cooling theories predict, but not as strongly as PSR B1055$-$52.

The standard cooling scenario assumes that NSs are
cooling down via the modified Urca neutrino emission process and
any processes of enhanced neutrino emission are not allowed. This scenario is 
largely independent of the dense matter equation of state and
stars which cool in such a way can be regarded as ``standard
candles'' of NS cooling theory \citep{yakovlev2011MNRAS}. If
J1741 is actually 390 kyr old and has the effective
temperature of 7~$\times$~10$^{5}$ K, 
calculations show that the neutrino
emissivity inside the star has been suppressed with respect to the
standard level throughout its life (Fig.~\ref{fig:cool}),
approximately by a factor of $\sim$~4. 

The suppression of the neutrino emission can be realized in the
framework of the minimal cooling theory \citep{gusakov2004AsAp,
page2004ApJS} which includes effects of the nucleon superfluidity.
In this model, the modified Urca processes are 
suppressed by the strong proton superfluidity. 
However, the minimal cooling
also assumes the presence of the neutron superfluidity
which enhances the cooling with respect to the standard level
\citep[see e.g.,][for details]{gusakov2004AsAp, page2004ApJS}. This
enhancement is required to explain the coldest sources in
Fig.~\ref{fig:cool}, however it makes impossible to fit the data
for hot  sources utilizing the usual models for neutron
superfluidity \citep{page2004ApJS,page2009ApJ}. The sources  at the
``knee'', such as  PSR B1055$-$52 and now J1741, can be
reconciled with the minimal cooling if one shifts the neutron
superfluidity to high densities, so that it does not take place in
low-mass stars (with low central density) and operates in
high-mass stars \citep{gusakov2004AsAp}. This assumption makes it
possible to explain all current data on cooling isolated neutron
stars in the unified way, including the possibly real-time cooling
NS in supernova remnant Cassiopeia~A \citep{shternin2011MNRAS}. In
this model the J1741 should have a low mass while the cold
NSs should be massive. These considerations remain true
if J1741 is several times younger and/or colder.
In this case the J1741 position on the cooling plane
will agree with the standard cooling scenario. Nevertheless, the   %
presence of the superfluidity inside the NSs is almost undoubted
and the question remains only about its quantitative
characteristics.

It is indeed possible that further analysis will result in
lower age or surface temperature of J1741.
An NS characteristic age can be
several times larger than its true age \citep[see
e.g.,][]{brisken2003AJ,thorsett2003ApJ}. The lower temperature of
the stellar surface also can not be ruled out. For instance, the
UV observations of PSR B1055$-$52
show a Rayleigh--Jeans component which exceeds the extrapolation of the X-ray
thermal spectrum by a factor of 4. This led 
\citet{mignani2010ApJ} to suggest that the X-ray emission in fact  
comes from the smaller hot region on the stellar surface (this is
possible if the pulsar is closer), while the entire surface is 
colder, being invisible in X-rays and showing itself in UV.
For illustration, with the open diamond and the label 3' in
Fig.~\ref{fig:cool} we show the position of the PSR B1055$-$52
according to the UV observations by \citet{mignani2010ApJ} (for
the distance 350~pc). This makes the positions of PSRs B1055$-$52
and J1741 similar with respect to standard cooling curves.
Note, that for the other two musketeers an extrapolation of
X-ray thermal spectra agrees with the UV data, while for RX J1856.5$-$3754
and other NSs of the ``Magnificent seven'' the situation is similar to
that of PSR B1055$-$52 \citep{kaplan2011ApJ}. It is therefore interesting to investigate 
which is the case for the J1741 in the UV. 

To conclude, the X-ray spectra of J1741
can be well described by a sum of a power law and a single blackbody with
temperature of about 60 eV.
If the blackbody component is interpreted as a thermal emission
from the entire surface of the neutron star, the distance to the
pulsar should be approximately 0.8 kpc. Similar value follows from
the analysis of the extinction towards the pulsar, supporting
this interpretation. The J1741 is a rather hot
middle-aged neutron star and its further studies will be useful to constrain the
physical input of the modern cooling theories. 

\acknowledgments
This research has made use of data obtained from the Chandra Data Archive 
and the Chandra Source Catalog, and software provided by 
the Chandra X-ray Center (CXC) in the application packages CIAO, ChIPS, and Sherpa.
We thank the anonymous referee for useful comments, 
and Dmitrii Barsukov for helpful discussion. 
The work was partially supported by  
the Russian Foundation for Basic 
Research (grants 13-02-12017-ofi-m, 14-02-00868-a) and
RF Presidential Programme MK-2837.2014.2.

{\it Facilities:} \facility{\textit{Chandra}}.  

\bibliographystyle{apj}
\bibliography{ref1741}



\clearpage


\end{document}